# Two-frequency heating technique at the 18 GHz electron cyclotron resonance ion source of the National Institute of Radiological Sciences[a]

S. Biri,[1,2,b] A. Kitagawa,[1] M. Muramatsu,[1] A. G. Drentje,[1] R. Rácz,[2] K. Yano,[3] Y. Kato,[3] N. Sasaki,[4] and W. Takasugi[4]

[1]*National Institute of Radiological Science (NIRS), 4-9-1 Anagawa, Inage, Chiba 263-8555, Japan*
[2]*Institute for Nuclear Research (ATOMKI), H-4026 Debrecen, Bem tér 18/c, Hungary*
[3]*Graduated School of Engineering, Osaka University, 2-1 Yamada-oka, Suita-shi, Osaka 565-0871, Japan*
[4]*Accelerator Engineering Corporation (AEC), Inage, Chiba 263-0043, Japan*

The two-frequency heating technique was studied to increase the beam intensities of highly charged ions provided by the high-voltage extraction configuration (HEC) ion source at the National Institute of Radiological Sciences (NIRS). The observed dependences on microwave power and frequency suggested that this technique improved plasma stability but it required precise frequency tuning and more microwave power than was available before 2013. Recently, a new, high-power (1200 W) wide bandwidth (17.1–18.5 GHz) travelling-wave-tube amplifier (TWTA) was installed. After some single tests with klystron and TWT amplifiers the simultaneous injection of the two microwaves has been successfully realized. The dependence of highly charged ions (HCI) currents on the superposed microwave power was studied by changing only the output power of one of the two amplifiers, alternatively. While operating the klystron on its fixed 18.0 GHz, the frequency of the TWTA was swept within its full limits (17.1–18.5 GHz), and the effect of this frequency on the HCI-production rate was examined under several operation conditions. As an overall result, new beam records of highly charged argon, krypton, and xenon beams were obtained at the NIRS-HEC ion source by this high-power two-frequency operation mode.

## I. INTRODUCTION

In the Heavy Ion Medical Accelerator in Chiba (HIMAC) of the National Institute of Radiological Sciences (NIRS),[1] efforts to extend the range of ion species of the ECR ion sources are continuously devoted. One way is feeding RF power into an ECRIS at two frequencies. This technique was originally initiated by ECR pioneers Jongen and Lyneis in Berkeley and some years later more successfully by Xie and Lyneis.[2] Since then many ECR laboratories have tested this technique and some of them use it in the everyday operation. The right choice of even the main frequency is not easy for a given ECRIS. Many reports pointed the importance of fine tuning of the (first and/or second microwave) frequency, for example, in Ref. 3. A detailed study on frequency tuning effects was performed by the Italian Istituto Nazionale di Fisica Nucleare (INFN, Italy), Gesellschaft für Schwerionenforschung (GSI, Germany), Jyvaskylan Yliopisto Fysiikan Laitos (JYFL, Finland) group.[4–7] Here the emphasis in the explanation was set to the RF properties of the plasma chamber. One of the present authors found similar results for more simple system.[8] The two-frequency heating technique has several advantages: it is effective for any kinds of ion species, no major modifica-

In early stages of our developments at NIRS, the enhancement of plasma region at different ECR zones was observed by the shapes of visible radiations.[9] The output currents had strongly depended on the additional microwave frequency.[10] Between 1998 and 2012 numerous two-frequencies experiments were carried out in NIRS. While the mechanism is still not clear, the positive effect of the second microwave was proved in each experiment. In spite of the limited maximum power and bandwidth of the additional microwave source, both the ion choice and beam intensities increased year by year.[11]

It became clear however, that in order to obtain more improvement, a larger microwave amplifier with enough wide bandwidth is necessary. Since from our experiences a bandwidth of a few percent is required, a klystron amplifier is not suitable as an additional microwave amplifier. A travelling wave tube amplifier is the optimal solution of two frequency heating technique at present. In the situation of the high-voltage extraction configuration (HEC) ion source of NIRS (called NIRS-HEC) the existing 700 W system was not enough for most cases. Therefore, recently a high-power second microwave system was developed, installed, and tested. The results of the first tests are presented.

## II. EXPERIMENTAL SETUP

An 18 GHz room-temperature ECR ion source, named NIRS-HEC, has been operating at the HIMAC accelerator of NIRS since 1996. The detailed technical description of NIRS-HEC is published in Ref. 12.

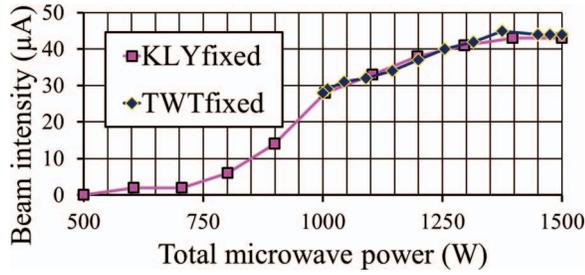

FIG. 1. Dependence of the Ar$^{13+}$ beam intensity on microwave power.

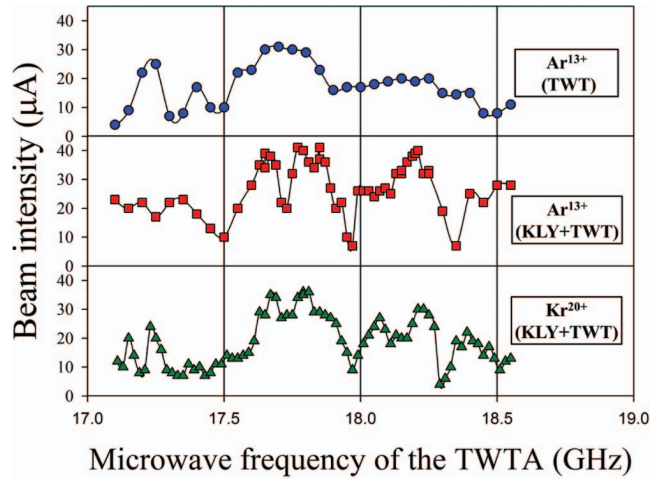

FIG. 2. Dependence of the beam intensity on the TWT frequency.

It is equipped with a klystron amplifier system (KLY) with a maximum power of 1500 W. In 2013 a travelling wave tube (TWT) amplifier system with frequency range from 17.10 to 18.55 GHz and with maximum power of 1200 W was added to NIRS-HEC, as second microwave source. Based on our earlier observations the coupling efficiency of the TWT power is near 1, while this value is only about 0.5 for KLY. These coefficients are considered to express the strength of coupling between the microwave and the plasma and are used later in Figure 1 and also in the rest of the paper. The technical realization of the TWT system is very similar to the one published in Ref. 11, except that now two 700 W TWT units are combined into one. The resulted total power would be 1400 W if the transmission coefficient from the synthesizer to the plasma chamber is 100%. In our practice the maximum output power measured in the waveguide close to the plasma chamber is 1200 W only. The TWT waveguide inside the ECRIS is water-cooled, similarly to the KLY waveguide, to the biased disk and to the plasma chamber. We note that NIRS-HEC cooling is very good, the outgassing is minimal below 2.5 kW total microwave power.

## III. DEPENDENCE ON MICROWAVE POWER

In an earlier paper,[11] we summarized the effect of microwave power to the stability of the extracted beam current. Our basic observation is that when the KLY power increases, the plasma shows instability and it is difficult to keep. When an additional microwave is added in the above situation, the plasma instability is improved at larger microwave power obtained by the mixture of two different frequency microwaves.

Figure 1 shows the dependence of the beam intensity of Ar$^{13+}$ on microwave power by operating NIRS-HEC at the new two-frequencies mode by changing the output power of one of the amplifiers, keeping the other power fixed. At this measurement all operation parameters were optimized by the mixed KLY and TWT: mirror magnetic fields, gas flows of Ar (main gas) and O$_2$ (mixing gas), extraction voltage, distance of puller, biased disk voltage, and the frequency of TWT. Then the microwave powers from both amplifiers were slightly varied again. The present measurement procedure was not same as the previous ones. One curve shows the case when the KLY power was fixed (1000 W) and the TWT power was varied from zero to 1000 W. In case of the other curve the TWT power was fixed (1000 W), while the KLY power was varied from zero to 1000 W. Using the mentioned coefficients the two curves draw are very similar.

## IV. DEPENDENCE ON MICROWAVE FREQUENCY

Figure 2 shows the dependence of the beam intensity of Ar$^{13+}$ on microwave frequency in the same basic condition as Figure 1. The dependence on the frequency of single microwave by TWT makes a peak around 17.7 GHz. The result suggests the source was well optimized at 17.7 GHz. The dependence of the Ar$^{13+}$ beam on the frequency of additional microwave by TWT has several peaks. The curves are partly similar, partly different, and it is very difficult to draw a clear prompt conclusion from them. We measured this frequency dependence several times. In most cases the beam current showed a peak (or at least a local peak) around TWT frequencies of 17.8 GHz and 18.2 GHz. The beam current here is always higher than setting TWT close or to exactly to 18 GHz. Figure 3 also shows a frequency dependence of Kr$^{20+}$. Although the condition was independently optimized, the tendency of frequency dependence was similar as the case of Ar$^{13+}$.

These observations suggest that an electron orbit effect might play some role. If the two frequencies differ from each other by only 200 MHz then the distance between the resonance layers is 0.7–0.9 mm in case of NIRS-HEC. The diameter of the Langmuir-orbit of a 5 keV electron is 0.74 mm, while it is 1.05 mm for the case of a 10 keV electron. These electrons thus can penetrate two resonance surfaces by one orbital turn and can receive an influence for the better plasma stability. (We are not sure at this moment that higher average electron energy or the better trapping of these electrons is the dominant process for a better plasma stability.) To verify or to reject this assumption a simple computer simulation was carried out by the TrapCAD code.[13] The same code was used successfully to optimize the magnetic trap of NIRS-HEC.[14] The geometry and the magnetic structure (using typical coils currents) of NIRS-HEC were used as input data. The code was run in the two-frequency mode, i.e., beside a fixed 18 GHz frequency a second resonance zone was applied. The frequency of this second microwave was varied between 17 and 19 GHz by 50 MHz steps. For simplicity both microwave powers were set to 1000 W. The 10 000 electrons were started from the 18 GHz resonance surface with fixed parallel energy (100 eV) and with random perpendicular energy between 100 eV and 10 000 eV. After 200 ns simulation time the average energies of the non-lost (plasma) electrons were saved. The electrons energy curve shows two local maxima around 17.75 and 18.2 GHz (Figure 3, upper curve). For reference we repeated the whole calculation at single frequency (varied between 17–19 GHz, microwave power 2 kW) without getting any peaks. The two curves meet at 18 GHz verifying the two-frequency simulation gives correct results. The figure also shows the number of the non-lost electrons.

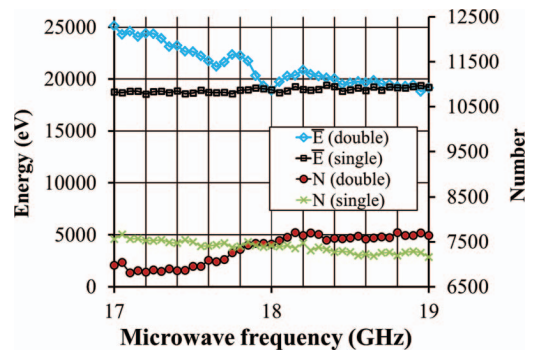

FIG. 3. Electron numbers and average energies vs microwave frequency. TrapCAD simulation.

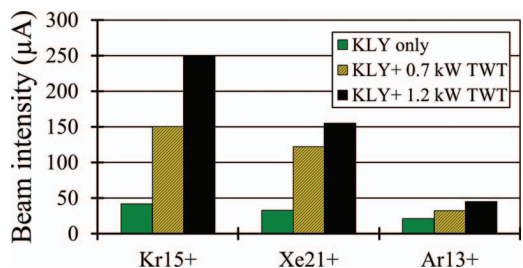

FIG. 4. Progress of selected beam intensities on the applied microwave injection mode.

These results are at least in coincidence with most of our experimental observations and suggest to continue a deeper analysis of the simulated results (comparisons of the electrons density distributions, power dependence, studying the trapping rates, size of the plasma, etc.) and also to carry out more experiments under other circumstances.

## V. PROGRESS FOR THE HCI RATE

After obtaining this experience with the basic features of the two microwave injection (power, TWT frequency) we examined how the production rate of the highly charged ions increases. NIRS-HEC was carefully tuned (by a time-consuming way) to the highest beam currents of one selected charge state of argon, krypton, and xenon. The results are shown in Figure 4 together with the previous beam current records for the same ions, obtained in earlier years at lower microwave powers. In each case significant increase of beam currents were achieved. We note that even when the beam current was saturated and there was no way to further increase it, the stability of this current and all the parameters of the ion source were very satisfying.

## VI. SUMMARY

The presented work showed that the combination of high-power KLY and high-power TWT amplifiers proved to be successful for NIRS-HEC. The stability of the plasma is better than just applying one single high-power frequency (regardless KLY or TWT). The output currents of the tested highly charged ions increased. To find the optimal second (TWT) frequency is still under considerations. Our experiment showed that either two substantially differing frequencies (e.g., 18 + 14 GHz) or two identical 18 GHz frequencies are performing worse than the application of two discrete, but close frequencies. For NIRS-HEC a 200–250 MHz frequency difference proved to be the best in these measurements.

[1] A. Kitagawa et al., Rev. Sci. Instrum. **83**, 02A332 (2012).
[2] Z. Q. Xie and C. M. Lyneis, in *Proceedings of the 12th International Workshop on ECRIS, Wako*, INS-J-182 (Institute for Nuclear Study, University of Tokyo, Japan, 1995), p. 24.
[3] P. Sortais et al., in *Proceedings of the 12th International Workshop on ECRIS, Wako*, INS-J-182 (Institute for Nuclear Study, University of Tokyo, Japan, 1995), p. 44.
[4] L. Celona et al., Rev. Sci. Instrum. **79**, 023305 (2008).
[5] V. Toivanen, H. Koivisto, O. Steczkiewicz, L. Celona, O. Tarvainen, T. Ropponen, S. Gammino, D. Mascali, and G. Ciavola, Rev. Sci. Instrum. **81**, 02A319 (2010).
[6] D. Mascali, L. Neri, S. Gammino, L. Celona, G. Ciavola, N. Gambino, R. Miracoli, and S. Chikin, Rev. Sci. Instrum. **81**, 02A334 (2010).
[7] F. Maimone, L. Celona, R. Lang, J. Mäder, J. Roßbach, P. Spädtke, and K. Tinschert, Rev. Sci. Instrum. **82**, 123302 (2011).
[8] Y. Kato, H. Furuki, T. Asaji, and S. Ishii, Rev. Sci. Instrum. **75**, 1470 (2004).
[9] A. Kitagawa et al., Rev. Sci. Instrum. **71**, 1061 (2000).
[10] A. Kitagawa et al., in *Proceedings of EPAC*, Wien (European Physical Society, Interdivisional Group on Accelerators, 2000), p. 1607.
[11] A. Kitagawa, M. Muramatsu, S. Biri, A. G. Drentje, W. Takasugi, and M. Wakaisami, in *Proceedings of the 18th International Workshop on ECRIS*, Chicago (Argonne National Laboratory, Chicago, USA, 2008), p. 92, see http://accelconf.web.cern.ch/AccelConf/ecris08/papers/mopo-12.pdf.
[12] A. Kitagawa, M. Muramatsu, M. Sekiguchi, S. Yamada, T. Okada, M. Yamamoto, S. Biri, and K. Uno, in *Proceedings of the 14th International Workshop on ECRIS*, Geneva (European Organization for Nuclear Research, CERN, PS Division, Switzerland, 1999), p. 23.
[13] S. Biri, R. Rácz, J. Imrek, A. Derzsi, and Zs. Lécz, IEEE Trans. Plasma Sci. **39**, 2474 (2011).
[14] A. Kitagawa, M. Muramatsu, M. Sekiguchi, S. Yamada, K. Jincho, T. Okada, M. Yamamoto, S. Biri, and K. Uno, Rev. Sci. Instrum. **71**, 981 (2000).